\documentstyle[12pt]{article}
\newcommand{\be}{\begin{equation}}
\newcommand{\ee}{\end{equation}}
\newcommand{\bea}{\begin{eqnarray}}
\newcommand{\eea}{\end{eqnarray}}
\newcommand{\norsl}{\normalsize\sl}
\newcommand{\norsc}{\normalsize\sc}

\def\Av{\mbox{\boldmath $A$}}

\def\Cv{\mbox{\boldmath $C$}}

\def\Kv{\mbox{\boldmath $K$}}

\def\qv{\mbox{\boldmath $q$}}
\def\kv{\mbox{\boldmath $k$}}

\begin{document}

\begin{titlepage}

\title{ Polarized Virtual Photon Structure Function
\\  in the Next-to-leading Order in  QCD}
\author{
\norsc  Ken SASAKI\thanks{e-mail address: sasaki@ed.ynu.ac.jp}~ and
      Tsuneo UEMATSU\thanks{e-mail address: uematsu@phys.h.kyoto-u.ac.jp} \\
\norsl  $^*$Dept. of Physics,  Faculty of Engineering, Yokohama National
University \\
\norsl  Yokohama 240-8501, JAPAN \\
\norsl  $^{\dag}$Dept. of Fundamental Sciences, FIHS, Kyoto University \\
\norsl     Kyoto 606-8501, JAPAN \\
}

\date{}
\maketitle

\begin{abstract}
{\normalsize The virtual photon structure function
$g_1^{\gamma}(x,Q^2,P^2)$, which can be
obtained in polarized $e^{+}e^{-}$ colliding-beam experiments, is investigated
for $\Lambda^2 \ll P^2 \ll Q^2$, where $-Q^2$ ($-P^2$) is the mass squared of
the probe (target) photon. The analysis is made to next-to-leading order
in QCD, in the framework of the QCD improved parton model with the DGLAP
evolution equations.  The non-leading corrections significantly modify the
leading log result,  in particular, at large $x$ as well as at small $x$.
}
\end{abstract}

\begin{picture}(5,2)(-290,-500)
\put(2.3,-115){YNU-HEPTh-98-102}
\put(2.3,-130){KUCP-124}
\put(2.3,-145){December 1998}
\end{picture}

\thispagestyle{empty}
\end{titlepage}
\setcounter{page}{1}
\baselineskip 18pt

In the last several years, the nulcleon spin structure functions
$g_1^{p(n)}(x,Q^2)$ and $g_2^{p(n)}(x,Q^2)$ have been extensively
studied by deep-inelastic scattering of polarized leptons on polarized
nucleon targets. The information on the spin structure of photon
would be provided by the resolved-photon process in polarized electron and
proton
collision in the polarized version of HERA\cite{Barber}. More directly, the
spin-dependent  structure function of photon can be measured by the
polarized  e$^+$ e$^{-}$ collision in the future linear colliders.
For real photon ($P^2=0$) target, there exists only one spin struture function,
$g_1^\gamma(x,Q^2)$, which is equivalent to the structure function $W_4^\gamma
(x,Q^2)$  ($g_1^\gamma\equiv 2W_4^\gamma$) discussed some time
ago in \cite{BM}.  The leading order (LO) QCD correction to $g_1^\gamma$
was first studied in \cite{KS} and later in \cite{MANO,BASS}.
Recently the first moment of
$g_1^\gamma$ attracted attention in the literatures~\cite{BASS,ET,NSV} in
connection  with its relevance to the axial anomaly, which also plays an
important  role in the QCD analysis of the nulcleon spin structure function
$g_1^{p(n)}$.

Now the next-to-leading order (NLO)
QCD analysis for $g_1^\gamma$ is possible, since the required spin-dependent
two-loop splitting functions in DGLAP evolution equations or,
equivalently,  the two-loop anomalous dimensions of the relevant operators
have been
calculated recently~\cite{MvN,V}. Actually there
has  been already an analysis of $g_1^\gamma$
for the case of the real photon target by Stratmann and Vogelsang \cite{SV}.

In this paper we shall investigate the polarized virtual photon structure
function $g_1^\gamma(x,Q^2,P^2)$ to the NLO in QCD, where $-Q^2$ ($-P^2$)
is the mass
squared  of the probe (target) photon (Fig.1).
We examine $g_1^\gamma(x,Q^2,P^2)$
for the kinematical region: $\Lambda^2 \ll P^2 \ll Q^2$~, where $\Lambda$
is the QCD scale parameter.
The advantage in studying the virtual photon target is that we can calculate
the whole structure function entirely up to NLO by the perturbative method.
On the
other hand, for the real photon target there appear  the non-perturbative
pieces in
NLO \cite{BB}, which we may dispose of by using, say,  vector-meson-dominance 
model. The unpolarized virtual photon structure functions
$F_2^\gamma(x,Q^2,P^2)$, and $F_L^\gamma(x,Q^2,P^2)$ were studied in LO
\cite{UW1} and in NLO \cite{UW2}.

The analysis can be performed either in the framework of the QCD improved
parton model \cite{A}
using the DGLAP parton evolution equations or in the framework
of the operator product expansion (OPE) supplemented by the renormalization
group
(RG) method.  In this paper we follow the former approach.

Let $q^i_{\pm}(x,Q^2,P^2)$, $G^{\gamma}_{\pm}(x,Q^2,P^2)$,
$\Gamma^{\gamma}_{\pm}(x,Q^2,P^2)$ be  quark with
$i$-flavor, gluon, and  photon distribution functions with $\pm$ helicities
of the longitudinally polarized virtual photon with mass $-P^2$.
Then the spin-dependent parton distributions are defined as
\bea
    \Delta q^i &\equiv& q^i_+ + \bar{q}^i_+ -q^i_- -\bar{q}^i_- \\
  \Delta G^{\gamma} &\equiv& G^{\gamma}_+ -G^{\gamma}_- ~,
      \qquad \Delta \Gamma^{\gamma} \equiv \Gamma^{\gamma}_+
-\Gamma^{\gamma}_- ~.
\eea
In the leading order of the electromagnetic coupling constant,
$\alpha=e^2/4\pi$,
$\Delta \Gamma^{\gamma}$ does not evolve with $Q^2$ and is set to be
$\Delta \Gamma^{\gamma}(x,Q^2,P^2)=\delta(1-x)$. For later convenience we use,
instead of $\Delta q^i$, the flavor singlet and non-singlet combinations of
spin-dependent quark distributions as follows:
\bea
     \Delta q^{\gamma}_S &\equiv& \sum_i~ \Delta q^i  \\
   \Delta q^{\gamma}_{NS} &\equiv& \sum_i e_i^2 \Bigl( \Delta q^i - \frac{\Delta
q^{\gamma}_S}{n_f} \Bigr)~,
\eea
where $n_f$ is the number of flavors of active quarks.
In terms of these parton distribution functions, the polarized virtual photon
structure function $g_1^{\gamma}(x, Q^2, P^2)$ is expressed
in the QCD improved parton model as\footnote{The detailed derivation of
this formula
will be reported elsewhere~\cite{SU}.}
\bea
   g_1^{\gamma}(x, Q^2,P^2)&=& \int^1_x \frac{dy}{y}~\biggl\{
   \Delta q^{\gamma}_S (y,Q^2,P^2)~C^S (\frac{x}{y}, Q^2)
                 + \Delta G^{\gamma} (y,Q^2,P^2)~C^G (\frac{x}{y}, Q^2)~
 \nonumber  \\
  & &  \quad + \Delta q^{\gamma}_{NS}
(y,Q^2,P^2) ~C^{NS} (\frac{x}{y}, Q^2)
\biggr\}   + C^{\gamma} (x, Q^2) \label{Solg}~,
\eea
where $C^S$($C^{NS}$), $C^G$, and $C^{\gamma}$ are the coefficient functions
corresponding to singlet(non-singlet)-quark, gluon, and photon, respectively,
and are independent of $P^2$.

The parton distributions $\Delta q^{\gamma}_S$, $\Delta G^{\gamma}$, and $\Delta
q^{\gamma}_{NS}$ evolve with $Q^2$. Their evolution equations
are expressed in a compact matrix form, once we introduce
a row vector $\Delta\qv^\gamma=(\Delta q^\gamma_S,~\Delta G^\gamma,~
\Delta q^\gamma_{NS})$:
\be
\frac{d~\Delta\qv^\gamma(x,Q^2,P^2)}{d~ \ln Q^2}=
\Delta\kv(x,Q^2) +
\int_x^1\frac{dy}{y}\Delta\qv^\gamma(y,Q^2,P^2)\Delta P(\frac{x}{y},Q^2)
\label{evol}~,
\ee
where the elements of a row vector $\Delta\kv= (\Delta K_S,~\Delta
K_G,~ \Delta K_{NS})$ are polarized splitting functions from virtual-photon to
singlet-quark, gluon, and non-singlet-quark, respectively.
The $3\times 3$ matrix $\Delta P(z,Q^2)$ is written as
\bea
\Delta P(z,Q^2)=
\left(\matrix{\Delta P^S_{q q}(z,Q^2)&\Delta P_{G q}(z,Q^2)&0\cr
           \Delta P_{q G}(z,Q^2)&\Delta P_{G G}(z,Q^2)&0 \cr
    0&0&\Delta P^{NS}_{q q}(z,Q^2)\cr}\right)~.
\eea
with its element $\Delta P_{AB}$ representing a splitting function
of $B$-parton to $A$-parton.

Once we get the information on the coefficient functions in
Eq.(\ref{Solg}) and
parton splitting functions in Eq.(\ref{evol}), we can
predict the behaviour of $g_1^{\gamma}(x, Q^2,P^2)$ in QCD. The NLO analysis is
now possible since the spin-dependent one-loop coefficient functions
and two-loop parton splitting functions are available~\cite{MvN,V}.
There are two methods to obtain $g_1^{\gamma}(x, Q^2,P^2)$ in NLO.
In one method, we use the parton splitting functions up to two-loop level and
we solve numerically $\Delta\qv^\gamma(x,Q^2,P^2)$ in Eq.(\ref{evol})
by iteration, starting
from the initial quark and gluon  distributions in the virtual photon at
$Q^2=P^2$. The interesting point of studying the virtual photon with mass
$-P^2$ is that  when $P^2 \gg \Lambda ^2$, the initial parton distributions
of the photon are completely known up to  NLO in QCD.
Then inserting the solved $\Delta\qv^\gamma(x,Q^2,P^2)$ into Eq.(\ref{Solg}),
and together with the known one-loop coefficient functions we can predict
$g_1^{\gamma}(x, Q^2,P^2)$ in NLO. 

The other method, which is more common than the former, is by making use of
the inverse Mellin transformation. We follow the latter method in this paper.
First we take the Mellin moments of Eq.(\ref{Solg}),
\be
     \int_0^1 dx x^{n-1}  g_1^{\gamma}(x,Q^2,P^2)
      = \Delta\qv^\gamma(n,Q^2,P^2) \cdot \Cv(n, Q^2) + C^{\gamma}_n
\label{Gmomento}
\ee
where we have defined the moments of an arbitrary function $f(x)$ as
\be
            f(n) \equiv \int_0^1 dx x^{n-1} f(x).
\ee
Henceforce we omit the obvious $n$-dependence for simplicity.
We expand the splitting functions $\Delta\kv(Q^2)$ and $\Delta P(Q^2)$
in powers of the QCD and QED coupling constants as
\bea
\Delta\kv(Q^2)&=&\frac{\alpha}{2\pi}\Delta\kv^{(0)}+
\frac{\alpha\alpha_s(Q^2)}{(2\pi)^2}\Delta\kv^{(1)} +\cdots \\
\Delta P(Q^2)&=&\frac{\alpha_s(Q^2)}{2\pi}\Delta P^{(0)}
+\left[\frac{\alpha_s(Q^2)}{2\pi}\right]^2 \Delta P^{(1)}
+\cdots  ~,
\eea
and introduce $t$ instead of $Q^2$ as the evolution variable~\cite{FP1},
\be
t \equiv \frac{2}{\beta_0}\ln\frac{\alpha_s(P^2)}{\alpha_s(Q^2)}~.
\ee
Then, taking the Mellin moments of the both sides in Eq.(\ref{evol}), we find
that ${\Delta\qv}^{\gamma}(t)(=\Delta\qv^\gamma(n,Q^2,P^2))$
satisfies the following inhomogenious differential equation \cite{GR}:
\bea
  \frac{d {\Delta\qv}^{\gamma}(t)}{d t}
 &=& \frac{\alpha}{2\pi} \Biggl\{ \frac{2\pi}{\alpha_s}{\Delta\kv}^{(0)}+
   \Bigl[ {\Delta\kv}^{(1)} - \frac{\beta_1}{2\beta_0}{\Delta\kv}^{(0)} \Bigr] +
   {\cal O}(\alpha_s) \Biggr\}  \nonumber \\
& &+ {\Delta\qv}^{\gamma}(t) \Biggl\{ \Delta P^{(0)}+ \frac{\alpha_s}{2\pi}
     \Bigl[ \Delta P^{(1)} - \frac{\beta_1}{2\beta_0}\Delta P^{(0)} \Bigr]  +
              {\cal O}(\alpha_s^2) \Biggr\}
\label{APE2}
\eea
where we have used the fact the QCD effective coupling constant
$\alpha_s(Q^2)$ satisfies
\be
  \frac{d \alpha_s(Q^2)}{d {\rm ln} Q^2}= -\beta_0 \frac{\alpha_s(Q^2)^2}{4\pi}
   -\beta_1 \frac{\alpha_s(Q^2)^3}{(4\pi)^2}+ \cdots
\label{beta}
\ee
with $\beta_0=11-2n_f/3$ and $\beta_1=102-38n_f/3$.
Note that the $P^2$ dependence of  $\Delta\qv^\gamma$ solely comes from the
initial condition (or boundary condition) as we will see below.

We look for the solution $\Delta\qv^\gamma(t)$ in the following form:
\be
\Delta\qv^\gamma(t)=\Delta\qv^{\gamma(0)}(t)+\Delta\qv^{\gamma(1)}(t)
\ee
where the first (second) term represents the solution in the LO (NLO).
First we discuss about the initial condition of $\Delta\qv^\gamma$. In the
framework of OPE, the twist-2 operators
$R_n^i$ ($i=S,G,NS,\gamma$) are relevant for the polarized photon structure
function $g_1^{\gamma}$. The expressions of $R_n^i$
are found, for example, in Ref.\cite{KS}. For $ -p^2 \equiv P^2 \gg \Lambda^2 $
we can calculate perturbatively the photon matrix
elements of the hadronic operators  $R_n^i$ ($i=S,G,NS$).
Choosing the renormalization point $\mu^2$ to be close to
$P^2$, we get, to the lowest order~\cite{UW2},
\be
   \langle \gamma (p) \mid R_n^i (\mu) \mid \gamma (p) \rangle =
\frac{\alpha}{4\pi}
   \Bigl( -\frac{1}{2}K_{n}^{0,i}~ {\rm ln}\frac{P^2}{\mu^2}+ A_{n}^i
\Bigr),
  \qquad  i=S,G,NS
\ee
where $K_{n}^{0,i}=({\Kv}^0_n)^i$ are one-loop  anomalous dimension matrices
between the photon and hadronic operators. In fact the $K_{n}^{0,i}$-terms
and $A_{n}^i$-terms represent the operator mixing between the photon operators
and  hadronic operators in  LO and NLO, respectively,
and the operator mixing implies that there exists quark and gluon
distributions in the
photon.   When we renormalize the photon matrix elements of the hadronic
operators at $\mu^2= P^2 $, we obtain
\be
  \langle \gamma (p) \mid R_n^i (\mu) \mid \gamma (p) \rangle \vert_{\mu^2= P^2}
=
     \frac{\alpha}{4\pi}  A_{n}^i
\label{Initial}
\ee
which shows that, at $\mu^2= P^2$, quark distribution exists in the photon, not
in the LO  but in the NLO.  Thus we have
\be
  {\Delta\qv}^{\gamma(0)}(0)=0, \qquad \quad
  {\Delta\qv}^{\gamma(1)}(0)= \frac{\alpha}{4\pi} {\Av}_n~.
\ee
Explicitly, ${\Av}_n$ is given by
\be
     {\Av}_n=6~(<e^2>,~0,~<e^4>-<e^2>^2)~ A^n_{qG}
\ee
where $<e^2>=\sum_i e^2_i/n_f$ and $<e^4>=\sum_i e^4_i/n_f$, and
$A^n_{qG}$ is the
finite part of the gluon matrix element of the  flavor-singlet quark
operator, whose expression in the $\overline{\rm MS}$ scheme is given
by~\cite{MvN},
\be
   A^n_{qG}=2n_f\biggl\{ \frac{n-1}{n(n+1)}\sum_{j=1}^n \frac{1}{j}+
   \frac{4}{(n+1)^2}-\frac{1}{n^2}-\frac{1}{n} \biggr\} ~.
\ee

With these initial conditions, we obtain for the solution
${\Delta\qv}^{\gamma}(t)$ of Eq.(\ref{APE2}) to the NLO (see \cite{SU} for
details):
\bea
&& {\Delta\qv}^{\gamma}(t)/\Bigl[ \frac{\alpha}{8\pi \beta_0}\Bigr] =
   \frac{4\pi}{\alpha_s(t)}~{\Kv}^0_n~\sum_i P^n_i~
       \frac{1}{1+\frac{\lambda^n_i}{2\beta_0}}
   \Biggl\{1-\biggl[\frac{\alpha_s(t)}{\alpha_s(0)}
   \biggr]^{1+\frac{\lambda^n_i}{2\beta_0}}  \Biggr\} \hspace{3cm}\nonumber  \\
&&\hspace{0.5cm}+  \Biggl\{{\Kv}^1_n~\sum_i P^n_i~
\frac{1}{\lambda_i^n/2\beta_0} + \frac{\beta_1}{\beta_0}{\Kv}^0_n
~\sum_i P^n_i~\Bigl( 1 - \frac{1}{\lambda_i^n/2\beta_0} \Bigr)  \nonumber  \\
& & \hspace{1cm}-{\Kv}^0_n \sum_{j,i} \frac{P^n_j~ 
\widehat \gamma^{(1)}_n~P^n_i}
  {2\beta_0+\lambda^n_j-\lambda^n_i}~\frac{1}{\lambda^n_i/2\beta_0}~
   - 2\beta_0{\Av_n}~\sum_i P^n_i  \Biggr\}
\Biggl\{1-\biggl[\frac{\alpha_s(t)}{\alpha_s(0)}
   \biggr]^{\frac{\lambda^n_i}{2\beta_0}}  \Biggr\}\nonumber  \\
&+&  \Biggl\{{\Kv}^0_n\sum_{i,j}~\frac{ P^n_i~\widehat \gamma^{(1)}_n
~P^n_j}{2\beta_0+\lambda^n_i-\lambda_j^n}~
\frac{1}{1+\frac{\lambda^n_i}{2\beta_0}}
 - \frac{\beta_1}{\beta_0}{\Kv}^0_n~
\sum_i~P^n_i~\frac{\lambda^n_i/2\beta_0}{1+\frac{\lambda^n_i} {2\beta_0}}
\Biggr\} 
\Biggl\{1-\biggl[\frac{\alpha_s(t)}{\alpha_s(0)}
   \biggr]^{1+\frac{\lambda^n_i}{2\beta_0}}  \Biggr\} \nonumber  \\
  &+& 2\beta_0{\Av}_n~.
\eea
Here the moments of the splitting functions are related to the anomalous 
dimensions of operators as follows \cite{MvN,V}:
\bea
   \Delta P^{(0)}&=&-\frac{1}{4}\widehat \gamma^0_n~, \qquad \qquad
   \Delta P^{(1)}=-\frac{1}{8}\widehat \gamma^{(1)}_n  \\
   {\Delta\kv}^{(0)}&=&\frac{1}{4}{\Kv}^0_n, \qquad \qquad \ \
    {\Delta\kv}^{(1)}=\frac{1}{8}{\Kv}^1_n
\eea
and we have introduced the projection operators given by
\bea
     \Delta P^{(0)}&=&-\frac{1}{4}\widehat \gamma^0_n =-\frac{1}{4}\sum_{i=+,-,NS}
\lambda^n_i~P^n_i,     \qquad   i=+,-,NS  \\
   P^n_i~P^n_j&=&\cases{0 &$i\ne j$, \cr
                         P^n_i &$i=j$},    \qquad
  \sum_i  P^n_i ={\bf 1}
\eea
where $\lambda^n_i$ are the eigenvalues of the matrix $\widehat \gamma^0_n$.
In fact, the parton distributions $\Delta\qv^\gamma(t)$ , {\it do} depend on
the initial conditions $\Delta\qv^\gamma(0)=(\alpha/4\pi)\Av_n$ but 
the structure function $g_1^\gamma(x,Q^2,P^2)$ itself is independent
of $\Delta\qv^\gamma(0)$ in NLO in QCD, which will be discussed in detail
in \cite{SU}.

When the coefficient functions are given, up to one-loop level, by,
\bea
     C^S(n,Q^2) &=& <e^2>\biggl\{ 1 + \frac{\alpha_S (Q^2)}{4\pi}
      B_S^n \biggr\}  \label{Bsing}\\
   C^G(n,Q^2)&=&<e^2> \Bigl\{ 0 + \frac{\alpha_S (Q^2)}{4\pi}
           B_G^n \Bigr\} \label{BG} \\
 C^{NS}(n,Q^2) &=&  1 + \frac{\alpha_S (Q^2)}{4\pi}  B_{NS}^n \\
   C^{\gamma}(n,Q^2)&=&\frac{\alpha}{4\pi}3 n_f <e^4> B_{\gamma}^n~,
         \label{Bgamma}
\eea
then we obtain from Eq.(\ref{Gmomento})
the following formula for the moments of $g_1^\gamma(x,Q^2,P^2)$ in the NLO:
\bea
&&\int_0^1 dx x^{n-1}g_1^\gamma(x,Q^2,P^2)  \nonumber\\
&=&\frac{\alpha}{4\pi}\frac{1}{2\beta_0}
\left[\sum_{i=+,-,NS}{\widetilde P}^n_i\frac{1}{1+\lambda_i^n/2\beta_0}
\frac{4\pi}{\alpha_s(Q^2)}
\left\{1-\left(\frac{\alpha_s(Q^2)}{\alpha_s(P^2)}\right)
^{\lambda_i^n/2\beta_0+1}\right\}\right.\nonumber\\
&&\hspace{2cm}+\left.\sum_{i=+,-,NS}{\cal A}_i^n\left\{1-\left(
\frac{\alpha_s(Q^2)}
{\alpha_s(P^2)}\right)^{\lambda_i^n/2\beta_0}\right\}\right.\nonumber\\
&&\hspace{2cm}+\left.\sum_{i=+,-,NS}{\cal B}_i^n\left\{1-
\left(\frac{\alpha_s(Q^2)}
{\alpha_s(P^2)}\right)^{\lambda_i^n/2\beta_0+1}\right\} +{\cal C}^n
+{\cal O}(\alpha_s) \right]
\label{master}
\eea
where ${\widetilde P}^n_i$, ${\cal A}_i^n$,
${\cal B}_i^n$ and ${\cal C}^n$
are  given by
\bea
   {\widetilde P}^n_i&=&{\Kv}_n^0 P^n_i {\Cv}_n(1,0) \label{tildeP}\\
   {\cal A}_i^n &=& -{\Kv}_n^0 \sum_j \frac{P^n_j \hat \gamma_n^{(1)} P^n_i}
             {2\beta_0 + \lambda^n_j - \lambda^n_i } {\Cv}_n(1,0)
     \frac{1}{\lambda^n_i/2\beta_0}
            -{\Kv}_n^0 \frac{\beta_1}{\beta_0}  P^n_i {\Cv}_n(1,0)
      \frac{1-\lambda^n_i/2\beta_0}{\lambda^n_i/2\beta_0} \nonumber  \\
    & & + {\Kv}_n^1 P^n_i {\Cv}_n(1,0) \frac{1}{\lambda^n_i/2\beta_0}
      - 2\beta_0 {\Av}_n P^n_i {\Cv}_n(1,0)  \label{calA} \\
\nonumber \\
 {\cal B}_i^n &=& {\Kv}_n^0 \sum_j \frac{P^n_i \hat \gamma_n^{(1)} P^n_j}
             {2\beta_0 + \lambda^n_i - \lambda^n_j } {\Cv}_n(1,0)
     \frac{1}{1+\lambda^n_i/2\beta_0}
    + {\Kv}_n^0 P^n_i \pmatrix{ <e^2> B^n_S  \cr
      <e^2> B^n_G \cr  B^n_{NS} \cr}
    \frac{1}{1+\lambda^n_i/2\beta_0} \nonumber  \\
   & &  -{\Kv}_n^0 \frac{\beta_1}{\beta_0}  P^n_i {\Cv}_n(1,0)
      \frac{\lambda^n_i/2\beta_0}{1+\lambda^n_i/2\beta_0}  \label{calB}\\
\nonumber \\
   {\cal C}^n &=& 2\beta_0 \Bigl\{3 n_f <e^4> B^n_{\gamma} +
         {\Av}_n \cdot {\Cv}_n(1,0)  \Bigr\}~,  \label{calC}
\eea
and  the column vector ${\Cv}_n(1,0)=(<e^2>, 0,1)^{\rm T}$ represents
tree-level coefficient functions. The expressions of Eqs.(\ref{master})
and (\ref{tildeP})--(\ref{calC}) are actually the same in form 
as the ones obtained before by
one of the authors and Walsh for the case of the virtual photon structure
function $F_2^{\gamma}$~\cite{UW2}.
All the quantities necessary to evaluate ${\widetilde P}^n_i$,
${\cal A}_i^n$, ${\cal B}_i^n$, and ${\cal C}^n$  are now known, which will be
reported in~\cite{SU}.
Especially we have at hand the results of the two-loop anomalous
dimensions~\cite{MvN,V} and one-loop coefficient functions 
$B^n_l\  (l=S,G,NS,\gamma)$
in Eqs.(\ref{Bsing})-(\ref{Bgamma})~\cite{KMMSU,BQ,MvN,V}  which were
calculated in the
$\overline{\rm MS}$ scheme.

In the case of  real photon target, the polarized structure function
$g_1^\gamma(x,Q^2)$ satisfies a remarkable sum rule which holds true in all
order
of $\alpha$ and $\alpha_S$~\cite{BASS,NSV,BBS} ,
\be
\int_0^1g_1^\gamma(x,Q^2)dx=0.
\ee
Now we can ask what happens to the first moment of the virtual photon
structure function $g_1^\gamma(x,Q^2, P^2)$. This can be studied by taking
the $n
\rightarrow 1$ limit of Eq.(\ref{master}).
Note that we have the following eigenvalues for the one-loop
anomalous dimension matrix ${\widehat \gamma}^0_{n=1}$:
\be
\lambda_{+}^{n=1}=0, \quad \lambda_{-}^{n=1}=-2\beta_0 , \quad
\lambda_{NS}^{n=1}=0
\label{zero1}
\ee
Physically speaking, the zero eigenvalues
$\lambda_{+}^{n=1}=\lambda_{NS}^{n=1}=0$
correspond to the conservation of the axial-vector current at one-loop
order. The other eigenvalue
$\lambda_{-}^{n=1}=-2\beta_0$, which is negative, is rather an artifact
of continuation of the gluon anomalous dimension to $n=1$, since
no gauge-invariant twist-2 gluon operator exists for $n=1$.
However, in the QCD improved parton model approach, there is no reason why
the $n=1$ moment of the polarized gluon distribution should not be
considered \cite{AL}. In the $n \rightarrow 1$ limit,  we expect in 
Eq.(\ref{master})
\begin{eqnarray}
&&\widetilde{P}_i^n\frac{1}{1+\lambda_i^n/2\beta_0} \rightarrow 0 \qquad
(i=+,-,NS) \nonumber\\
&&{\cal A}_{+}^n \rightarrow \mbox{finite}, \quad
{\cal A}_{-}^n \rightarrow 0, \quad {\cal A}_{NS}^n \rightarrow \mbox{finite}
\nonumber\\ &&{\cal B}_{+}^n \rightarrow 0, \quad {\cal B}_{-}^n \rightarrow
\mbox{finite},
\quad {\cal B}_{NS}^n \rightarrow 0~.
\end{eqnarray}
However, ${\cal A}_{+}^n $, ${\cal A}_{NS}^n$, and ${\cal B}_{-}^n$
are multiplied by the following vanishing factors
\begin{equation}
\left\{1-\left(\frac{{\bar g}^2(Q^2)}
{{\bar g}^2(P^2)}\right)^{\lambda_{+}^n/2\beta_0}\right\}, \quad
\left\{1-\left(\frac{{\bar g}^2(Q^2)}
{{\bar g}^2(P^2)}\right)^{\lambda_{NS}^n/2\beta_0}\right\}, \quad
\left\{1-\left(\frac{{\bar g}^2(Q^2)}{{\bar g}^2(P^2)}\right)
^{\lambda_{-}^n/2\beta_0+1}\right\}~,  \label{vanishing}
\end{equation}
respectively, and thus the terms proportional to ${\widetilde P}^n_i$,
${\cal A}_i^n$, and ${\cal B}_i^n$ in Eq.(\ref{master})
all vanish in the $n=1$ limit. Note that these vanishing factors are specific
to the case of the virtual photon target, and that such factors do not appear
when the target is real photon.  Thus we find
\begin{equation}
\int_0^1dx g_1^\gamma(x,Q^2,P^2)=\frac{\alpha}{4\pi}\frac{1}{2\beta_0}
{\cal C}^{n=1} +{\cal O}(\alpha_s)~.
\end{equation}

Now due to the relation $B^n_{\gamma}=\frac{2}{n_f}B^n_{G}$, we have from
Eq.(\ref{calC})
\be
        {\cal C}^{n=1}=12\beta_0<e^4>(B_G^n+A^n_{qG})\vert_{n=1}~.
\ee
It is noted that the combination $(B_G^n+A^n_{qG})$ is
renormalization-scheme independent~\cite{BBDM}.  The results in the
$\overline{\rm MS}$ scheme~\cite{MvN,V,BQ} are
\be
B_G^{n=1}=0, \qquad A^{n=1}_{qG}=-2n_f
\ee
The same results have been obtained by
Kodaira \cite{JK} in the framework of OPE and RG method. He set $B_G^{n=1}=0$,
observing that there is no gauge-invariant twist-2 gluon operator for $n=1$ and
obtained $A^{n=1}_{qG}=-2n_f$ from the Adler-Bell-Jackiw anomaly.
In the end, we have for the sum rule of the virtual photon
structure function $g_1^\gamma$,
\be
\int_0^1 dx g_1^\gamma(x,Q^2,P^2)=-\frac{3\alpha}{\pi}\sum_{i=1}^{n_f}
e_i^4 +{\cal O}(\alpha_s),  \qquad {\rm for }\ \  \Lambda^2 \ll  P^2 \ll Q^2~.
\ee
Finally it should be pointed out that we can further pursue the QCD
corrections of
order $\alpha_s$ to the first moment of $g_1^\gamma$~\cite{NSV},
which will be discussed in detail in~\cite{SU}. Here we
just write down the result which turns out to be:
\bea
\int_0^1dx g_1^\gamma(x,Q^2,P^2)
&=&-\frac{3\alpha}{\pi}
\left[\sum_{i=1}^{n_f}e_i^4\left(1-\frac{\alpha_s(Q^2)}{\pi}\right)
\right.\nonumber\\
&-&\left.\frac{2}{\beta_0}(\sum_{i=1}^{n_f}e_i^2)^2\left(
\frac{\alpha_s(P^2)}{\pi}-\frac{\alpha_s(Q^2)}{\pi}\right)\right]
+{\cal O}(\alpha_s^2).
\eea
This result is perfectly in agreement with the one obtained by
Narison, Shore and Veneziano in ref.\cite{NSV}, apart from the
overall sign for the definition of $g_1^\gamma$.

The polarized virtual photon structure function $g_1^\gamma(x,Q^2,P^2)$ is
recovered
from the moments, Eq.(\ref{master}) by the inverse Mellin transformation.
In Fig.2 we have plotted, as an illustration, the result for
$n_f=3$, $Q^2=30$GeV$^2$, $P^2=1$GeV$^2$ and the QCD scale parameter
$\Lambda=0.2$GeV. The vertical axis corresponds to
\be
g_1^\gamma(x,Q^2,P^2)/\frac{3\alpha}{\pi}~n_f<e^4>\ln\frac{Q^2}{P^2}~.
\label{normalized}
\ee
Here we have shown three cases; the box (tree) diagram
contribution,
\be
g_1^{\gamma(Box)}(x,Q^2,P^2)=(2x-1)\frac{3\alpha}{\pi}~n_f<e^4>\ln\frac{Q^2}
{P^2}~,
\ee
the LO  and NLO results in QCD.
We observe that the NLO QCD corrections are significant at large
$x$ as well as at low $x$. Other examples with different $Q^2$ and
$P^2$ will be reported in the forthcoming paper \cite{SU}. We have not
seen any sizable change for the normalized structure
function (\ref{normalized}).

Now let us consider the real photon case $P^2=0$. The structure function can be
decomposed as
\be
g_1^\gamma(x,Q^2)=g_1^\gamma(x,Q^2)\vert_{\rm pert.} +
g_1^\gamma(x,Q^2)\vert_{\rm non-pert.}~.
\ee
The second term can only be computed by some non-perturbative method such
as lattice
QCD, or estimated by vector-meson-dominance model. The first term,
the point-like
piece, can be calculated in a perturbative method. Actually, it can  be
reproduced formally by setting $P^2=\Lambda^2$ in Eq.(\ref{master}).
In Fig.3, we have plotted the point-like piece of $g_1^\gamma$ of the real
photon.
The LO QCD result coincides with the previous result obtained by one of the
authors
in \cite{KS}. The NLO result is qualitatively consistent with the analysis
made by
Stratmann and Vogelsang \cite{SV}.

In the present paper, we have investigated polarized virtual photon
structure function $g_1^\gamma(x,Q^2,P^2)$ in the NLO in QCD for the
kinematical
region $\Lambda^2\ll P^2 \ll Q^2$. The analysis has been made
in the framework of the QCD improved
parton model with the DGLAP parton evolution equations.
We have shown that the behavior of $g_1^\gamma(x,Q^2,P^2)$ can be predicted
up to NLO in QCD without any free parameter. In fact the NLO QCD corrections
are significant at large $x$ and also at small $x$.
The first moment of $g_1^\gamma$ for the virtual photon is non-vanishing
in contrast to the case of real photon where the first moment vanishes.

\bigskip

\bigskip

\bigskip

\bigskip

\vspace{0.5cm}
\leftline{\large\bf Acknowledgement}
\vspace{0.5cm}

We thank G. Altarelli, J. Kodaira, G. Ridolfi, G. Veneziano
and W. Vogelsang for valuable discussions. Part of this work
was done while one of us (T.U.) was at CERN. He thanks the
CERN TH Division for its hospitality. This work is supported in part
 by the Monbusho Grand-in-Aid for Scientific Research 
No.(C)(2)-09640342 (K.S.) and No.(C)(2)-09640345 (T.U.).

\newpage

\newpage

\vspace{0.5cm}
\leftline{\large\bf Figure Caption}
\vspace{0.5cm}

\noindent
Figure 1

\noindent
Deep inelastic scattering on a virtual photon in polarized
e$^+$ e$^{-}$ collision, \\
 e$^+$ e$^{-}$ $\rightarrow$ e$^+$ e$^{-}$ + hadrons.
The mass squared  of the ``probe" (``target") photon  is $-Q^2$ ($-P^2$)
( $\Lambda^2 \ll P^2 \ll Q^2$ ).

\bigskip

\bigskip

\noindent
Figure 2

\noindent
The polarized virtual photon structure function $g_1^\gamma(x,Q^2,P^2)$ to
the next-to-leading order (NLO) in units of
$\frac{3\alpha}{\pi}~n_f<e^4>\ln\frac{Q^2}{P^2}$ for $Q^2=30$GeV$^2$,
$P^2=1$GeV$^2$
and the QCD scale parameter $\Lambda=0.2$GeV with $n_f=3$.
We also plot the leading order (LO) result (long-dashed line) and the box
diagram
contribution (short-dashed line).

\bigskip

\bigskip

\noindent
Figure 3

\noindent
The point-like piece of the polarized real photon structure function
$g_1^\gamma(x,Q^2)$ to  the next-to-leading order (NLO) in units of
$\frac{3\alpha}{\pi}~n_f<e^4>\ln\frac{Q^2}{\Lambda^2}$ for $Q^2=30$GeV$^2$,
 and the QCD scale parameter $\Lambda=0.2$GeV with $n_f=3$.
We also plot the leading order (LO) result (long-dashed line) and the box
diagram
contribution (short-dashed line).

\newpage
\pagestyle{empty}
\input epsf.sty
\begin{figure}
\centerline{
\epsfxsize=11cm
\epsfbox{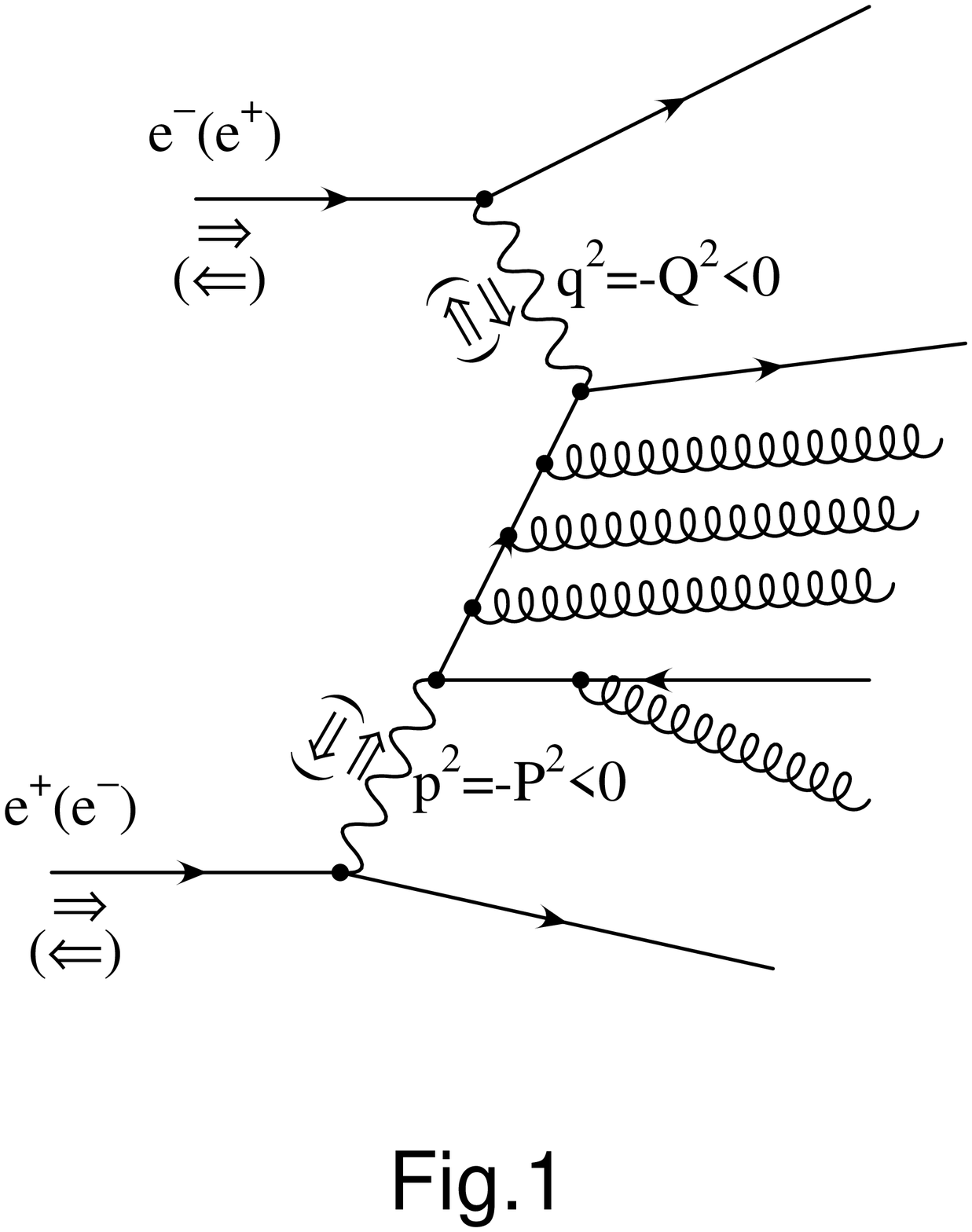}}
\end{figure}

\newpage
\pagestyle{empty}
\input epsf.sty
\begin{figure}
\centerline{
\epsfxsize=26cm
\epsfbox{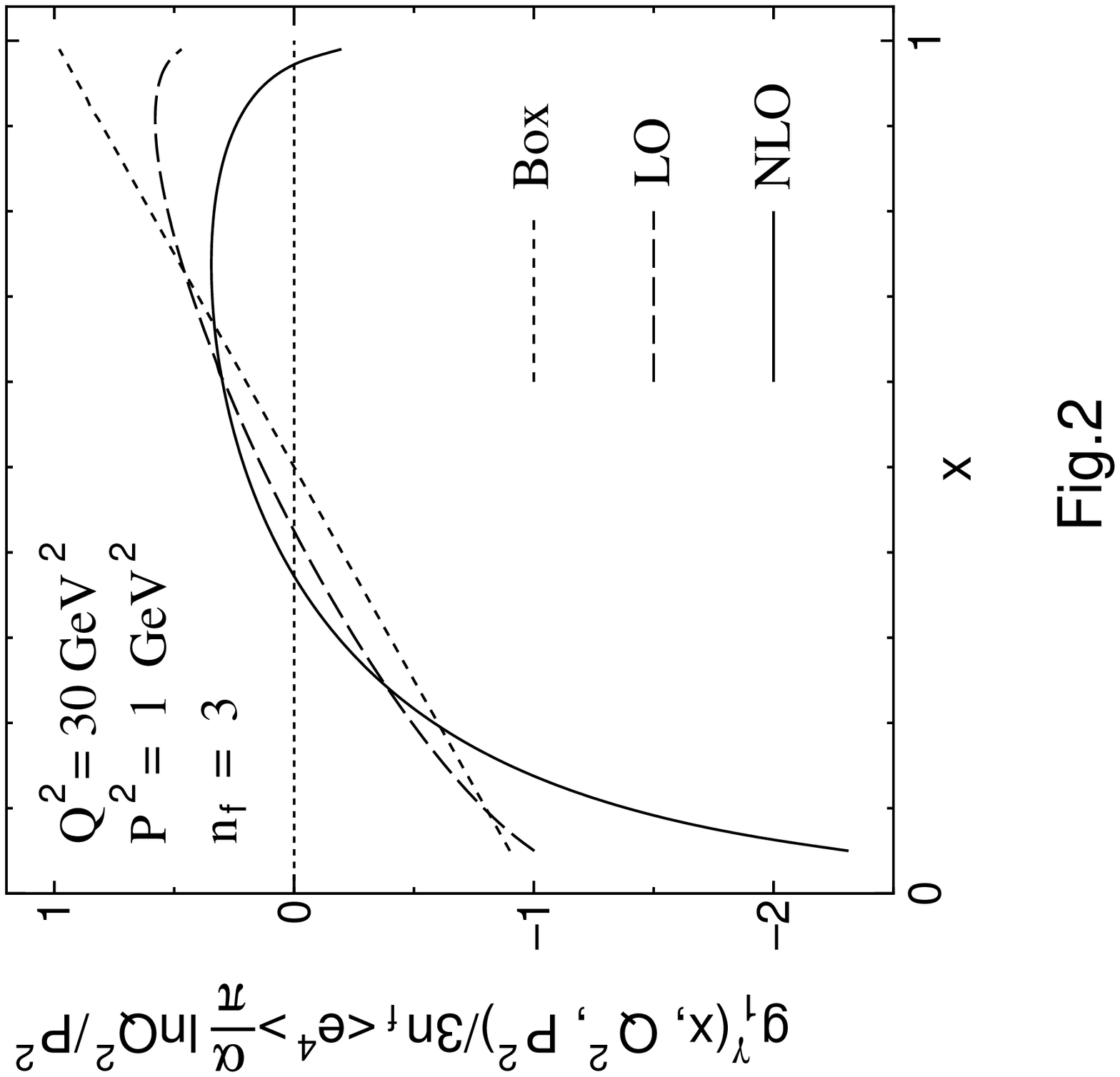}\hspace{2cm}}
\end{figure}

\newpage
\pagestyle{empty}
\input epsf.sty
\begin{figure}
\centerline{
\epsfxsize=26cm
\epsfbox{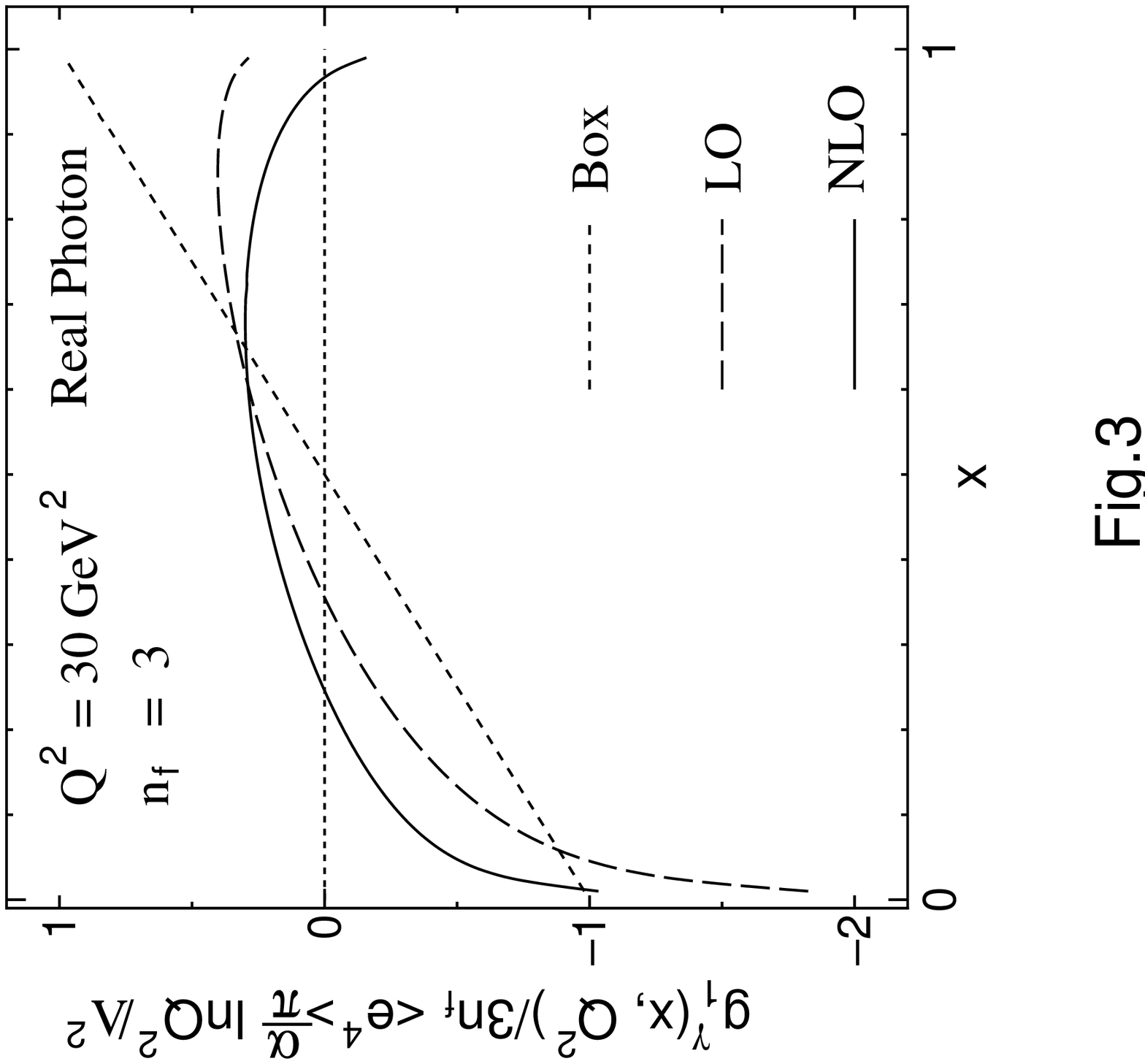}\hspace{2cm}}
\end{figure}

\end{document}